\documentclass[twocolumn]{aastex6}

\usepackage{amsmath}
\usepackage{natbib}
\usepackage{multirow}
\usepackage{graphicx}

\def\mwc{MWC~656}
\def\agl{AGL~J2241+4454}
\def\agile{\textit{AGILE}}
\def\fermi{\textit{Fermi}/LAT}
\newcommand{\grp}    {${\rlap.}^{\circ}$}

\begin{document}


\title{The gamma-ray source \agl\ as the possible counterpart of \mwc}


%
%
%
%
%

\author{Pere Munar-Adrover\altaffilmark{1}, S. Sabatini\altaffilmark{2}, Giovanni Piano and Marco Tavani\altaffilmark{2,3}} 
\affil{INAF/IAPS-Roma, I-00133 Roma, Italy}

\author{L. H. Nguyen}
\affil{Institut f\"ur Experimentalphysik, Universit\"at Hamburg, Germany}

\and 

\author{F. Lucarelli\altaffilmark{4}, F. Verrecchia\altaffilmark{4}, C. Pittori\altaffilmark{4}}
\affil{ASI Science Data Centre (ASDC), via del Politecnico snc, I-00133 Roma, Italy}


\altaffiltext{1}{pere.munar@iaps.inaf.it}
\altaffiltext{2}{INFN Roma Tor Vergata, I-00133 Roma, Italy}
\altaffiltext{3}{Dip. di Fisica, Univ. Tor Vergata, I-00133 Roma, Italy}
\altaffiltext{4}{INAF-OAR, via Frascati 33, I-00040 Monte Porzio Catone, Italy}

\begin{abstract}

The \agile\ satellite discovered the transient source \agl\ in 2010 which triggered the study of the associated field allowing for the discovery of the first Be/black hole binary system: \mwc. This binary was suggested to be the counterpart of \agl, but this is still not a firm association. In this work we explore the archival \agile\ and \fermi\ data in order to find more transient events originating in the field of \agl\ and address the possibility to link them to the accretion/ejection processes of \mwc. We found a total of 9 other transient events with \agile\ compatible with the position of \agl, besides the one from 2010. We folded this events with the period of the binary system and we did not find significant results that allow us to associate the gamma-ray activity to any particular orbital phase. By stacking the 10 transient events we obtained a spectrum that extends between 100~MeV and 1~GeV, and we fitted it with a power law with photon index $\Gamma =2.3\pm0.2$. We searched into the \fermi\ data in order to complement the gamma-ray information provided by \agile\ but no significant results arose. In order to investigate this apparent contradiction between the two gamma-ray telescopes, we studied the exposure of the field of \agl\ in both instruments finding significant differences. In particular, \agile\ exposed for longer time and at lower off-axis angle distance the field of \agl. This fact, together with the energy dependent sensitivity of both instruments, and the soft spectrum found in the stacking analysis, might explain why \agile\ observed the transient events not seen by \fermi.

\end{abstract}





\section{Introduction }


In July 2010 the \agile\ satellite detected a gamma-ray transient source, {\object AGL J2241+4454}, below the galactic plane \citep{Lucarelli10}, thus far from the strong diffuse Galactic emission. Within the \agile\ position error box there are only 4 prominent X-ray sources: {\object RX J2243.1+4441}, a radio quasar \citep{Brinkmann97}; \object{HD 215325}, an eclipsing binary of beta Lyr type \citep{kiraga2013}; \object{TYC 3226-1310-1}, a rotationally variable star \citep{kiraga2013}; and {\object MWC 656}, a massive Be star \citep{2007A&A...474..653V}. Both HD 215325 and TYC 3226-1310-1 star systems do not host the conditions for gamma-ray emission processes, such as accreting companions or the collision of two strong stellar winds, which could accelerate particles up to relativistic energies and thus produce gamma rays. As we will show throughout the paper, we consider \mwc\ as the most likely candidate to be the counterpart of the gamma-ray transient emission.

\mwc\ has been recently discovered as the first binary system containing a black hole (BH) and a Be companion star \citep{Casares14}. This system was discovered thanks to its putative association with the transient gamma-ray source \agl\ \citep{Lucarelli10} that triggered multiwavelength observations.

This binary system is located at galactic coordinates $(l, b)= ($100\grp18, -12\grp40$)$ and shows a photometric modulation of 60.37 $\pm$ 0.04 days \citep{Williams10, Paredes-Fortuny12} which suggested its binary nature, confirmed by \cite{Casares_optical_2012}.  Phase zero, $\phi_{0}$, is definded at the optical maximum (53242.8 MJD) and the periastron occurs at $\phi=0.01 \pm 0.1 $ \citep{Casares14}. In the latter paper, it was found that the compact object is a BH with a mass between 3.8--6.9 $M_{\odot}$, and the Be star was reclassified as a B1.5-B2 III type star with a mass between 10--16 $M_{\odot}$. X-ray follow up observations revealed a faint X-ray source \citep{munaradrover14} compatible with the position of the Be star, allowing for the classification of this system as a high-mass X-ray binary (HMXB). The X-ray spectrum was fitted with a black-body plus a powerlaw model. The total X-ray luminosity in the 0.3--5.5~keV band is $L_{\rm X}=(3.7\pm1.7)\times 10^{31}$ erg s$^{-1}$. The measured X-ray spectrum is dominated by the non-thermal component, with a non-thermal luminosity $L_{\rm pow}=(1.6^{+1.0}_{-0.9})\times 10^{31}$ erg s$^{-1} \equiv (3.1\pm2.3)\times 10^{-8} L_{\rm Edd}$ for the estimated BH mass. This very low X-ray luminosity is compatible with the binary system being in quiescence during the X-ray observations and is at the level of the faintest low-mass X-ray binaries (LMXBs) ever detected, such as A0620$-$00 \citep{Gallo2006} and XTE~J1118+480 \citep{Gallo2014}. The non-thermal component is interpreted in \cite{munaradrover14} as the contribution arising from the vicinity of the BH and is studied in the context of the accretion/ejection coupling seen in LMXBs \citep{Fender10}, and also in the context of the radio/X-ray luminosity correlation \citep{Corbel13, Gallo2014}.

Recent VLA radio observations by \cite{Dzib15} showed that the binary system is also a weak radio source with variable emission. The peak of the radio flux density is $14.2\pm2.9~\mu$Jy for a single $\sim$2h observation taken in 2015 February 22, at orbital phase $\sim$0.49, while 6 following radio observations carried out during 2015 at the same sensitivity did not show any radio signal. However, stacking these other 6 observations yielded a detection with a flux density of $3.7\pm1.4~\mu$Jy compatible with the position of \mwc.

At TeV energies, \mwc\ was observed in 2012 and in 2013 with the MAGIC Telescopes. The 2013 observatinons were contemporaneous to the {\it XMM-Newton} observation by \cite{munaradrover14}. According to X-ray and optical data, these observations were thus performed during an X-ray quiescent state of the binary system and yielded only upper limits to its possible very-high energy (VHE) emission at the level of $2.0\times10^{-12}$ cm$^{-2}$ s$^{-1}$ above 300~GeV \citep{aleksic2015}.

X-ray binary (XRB) systems containing a Be star have been studied for a long time through radio and X-ray surveys. These systems are characterized by their variable X-ray emission and for their strong radio flares. Out of 184 known BeXRB systems, in 119 of them the X-ray pulsations are observed, confirming that the compact component must be a neutron star \citep{Ziolkowski14}. In the remaining BeXRB, the nature of the compact object is still unknown, with no confirmed BH known with a Be companion. This is known as the missing Be/BH binary problem \citep{Belczynski07}, and the predicted number of existing systems of this kind at any time is a few tens in our Galaxy \citep{Grudzinska15}. In our case, the newly discovered \mwc\ system opens a new window in the study of this class of binaries. The missing Be/BH population could then possibly reveal itself through gamma-ray surveys more easily than with X-ray surveys and could have been subject to selection effects up to now.

In this work we present new \agile\ and \fermi\ data analysis: we searched for persistent, transient and periodic emission at the position of \mwc. We also discuss possible counterparts of the gamma-ray events and the possible association of the transient gamma-ray emission found with \mwc\ in the context of the accretion/ejection scenario.



\section{The \mwc\ field}\label{sec:field}

Although \mwc\ seems to be the most favoured candidate to produce the transient emission that has been detected by \agile, we cannot discard another source as the counterpart. 

In the vicinity of \agl\ there is a \fermi\ source, 3FGL~J2247.8+4413, which is identified with NVSS J224753+441317, a BL Lac object \citep{Fermi3fgl}. It is located at galactic coordinates $(l,b)=$ $($100\grp72,-13\grp25$)$, at a distance of 0\grp93 from the position of \agl. We considered this source at the time of analyzing \agile\ data in order to take into account possible contamination. However, this is a very faint source with a flux of $F(E>1 {\rm GeV} = (5.3\pm0.5)\times 10^{-10}$~cm$^{-2}$~s$^{-1}$ and variable behaviour. Hence, in the short time periods when \agile\ detected the transient emission the effect of 3FGL~J2247.8+4413 on the detection is negligible.

The error box of the first \agile\ transient detection is 1\grp2 in diameter. As for X-ray emitting sources in the field, there is a known quasar near \mwc, \object{RX J2243.1+4441} \citep{Brinkmann97}, which appears also in the X-ray images taken by \cite{munaradrover14}. Quasars are known to produce also flaring activity at many wavelengths \citep{Hartman2001} and this could be the case. However, there is little information about RX~J2243.1+4441, its behaviour and characteristics, due to the lack of optical counterpart and scarce radio and X-ray data. This source can be classified as a radio galaxy through its radio morphology \citep{Marcote_thesis}: it presents a bright core at the center of the galaxy with a radio flux of 2.2 mJy beam$^{-1}$, stable over month timescales with faint signatures of a jet, and two radio lobes expanding towards the intergalactic medium spanning more than 50$^{\prime\prime}$. However, the available radio data at only one frequency does not allow for a spectral characterization of the source. The radio morphology of this object resembles the classical FR-II type radio galaxies. While FR I-type radio galaxies have been found to emit in gamma rays in several cases \citep{Fermi3fgl}, FR-II type seem to be more rare in the GeV sky \citep{2010ApJ...720..912A, Grandi2012}, disfavouring, although not excluding, RX~J2243.1+4441 as the putative counterpart of \agl.

In the X-rays, a revision of the XMM-Newton data (Obs ID 0723610201), shows that the radio galaxy exhibits a spectrum with peak around 0.5 keV and fades away above 5 keV. The spectral fit with an absorbed powerlaw model yields a photon index $\Gamma=1.8\pm0.4$ and a flux in the 0.2$-$5.0 keV range of F$(0.2-5{\rm keV})=(6.0\pm0.7)\times10^{-14}$ erg cm$^{-2}$ s$^{-1}$. This results are roughly compatible within uncertainties with the \cite{Brinkmann97} results using {\it ROSAT}-PSPC data. For a detailed view of the \agl\ field see Appendix \ref{ap:field}

For the above reasons, in the rest of the paper we consider \mwc\ as the most probable candidate counterpart of \agl. 


\section{Observational data}
\subsection{ \agile\ data}


\agile\ is a gamma-ray astrophysics mission operating since 2007 April \citep{Tavani2008}. It carries two main co-aligned instruments observing at hard X-rays between 18 and 60 keV (Super-\agile\,, \citealt{Feroci07}) and at high-energy (HE) gamma rays between 30 MeV and 30 GeV (\agile /GRID, \citealt{Barbiellini2002, Prest2003}). The instrumentation is completed by a calorimeter sensitive in the 0.4--100 MeV range \citep{Labanti2006} and an anticoincidence detector \citep{Perotti06}. \agile\ has a wide field of view of about 2.5~sr at HE gamma rays and good sensitivity above 100 MeV. The PSF at 100 MeV and 400 MeV is 3\grp5 and 1\grp5 (68\% containment radius), respectively \citep{Sabatini15}. The sensitivity after one week of integration time in pointing mode is at the level of 20--30$\times 10^{-8}$ cm$^{-2}$ s$^{-1}$ above 100 MeV, depending on the off-axis angles and pointing directions \citep{Tavani2008}. 

Since November 2009 \agile\ is observing in scanning mode, covering the whole sky every few hours. This observation mode allows us for the study of serendipitous sources, such as AGL~J2241+4454. Typical sensitivity in this observing mode is $\sim 10^{-7}$ cm$^{-2}$ s$^{-1}$ for 2-day integration time.

\subsection{\textit{Fermi}/LAT data \label{sec:fermi}}

The \textit{Fermi} gamma-ray space telescope was launched in June 2008 and carries onboard two main instruments: the Large Area Telescope (LAT) and the Gamma-ray Burst Monitor (GBM). The LAT is a non-homogeneous pair conversion tracker that works in the 30 MeV to 300 GeV energy range and has a large collection area, ranging from 1000 to 5000 cm$^2$ for the front-LAT component. The instrument 68\% containment radius  ranges from 7$^{\rm \circ}$ to 0.1$^{\rm \circ}$ in the considered energy range and the instrument angular resolution is optimized around 1GeV \citep{Sabatini15}. \fermi\ operates in scanning mode, monotiring continuously a wide part of the sky every few hours, with a typical sensitivity of $10^{-7}$ cm$^{-2}$ s$^{-1}$ for 1-day integration. Note that, due to the scanning observing mode, sources spend a very small amount of time (if any) on-axis.

\section{Data analysis}

\subsection{\agile\ data analysis}
 The analysis of gamma-ray data presented in this work was carried out with the \agile/GRID FT3.119\_I0023 Build22 calibrated filter with a gamma-ray event selection that takes into account South Atlantic Anomaly event cuts and 80$^\circ$ Earth albedo filtering. Throughout the work, statistical significance assessment and source flux determination were established using the standard \agile\ multi-source likelihood analysis software \citep{Chen12}. The method provides an assessment of the statistical significance in terms of a Test Statistic (TS) defined as in \cite{Mattox96} and asymptotically distributed as a $\chi^2/2$ for 3 degrees of freedom ($\chi_3^2/2$).

\subsubsection{Search for gamma-ray persistent emission}

\begin{figure}[ht!]
        \includegraphics[width=\linewidth]{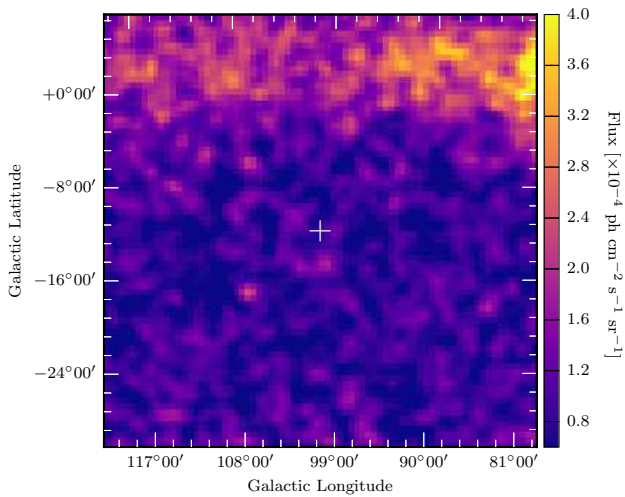}
        \caption{Deep \agile\ pointing integration intensity map around the position of \mwc\ (white cross). Color-map represents 3-pixel kernel gaussian smoothed number of counts. Pixel size is 0\grp2.\label{fig:deeppointing}}
\end{figure}

A search for persistent gamma-ray emission has been conducted over the public \agile\ consolidated archive available from the ASI Science Data Center\footnote{http://www.asdc.asi.it/mmia/index.php?mission=agilemmia} at the time of the analysis, which covered observations up to the end of 2013. We performed a multi-source likelyhood analysis above 100~MeV of the $\sim$7 years of \agile\ data and no detection arose so far and an upper limit at the level of $UL(E> 100 \rm{MeV}) = 2\times10^{-9}$ cm$^{-2}$ s$^{-1}$ has been obtained. We do not expect to find persistent gamma-ray emission in this system since XRBs spend most of their time in a quiescent state \citep{Plotkin13}. During this state there is no presence of a radio jet, or it is not powerful enough to accelerate particles up to HE, and hence the possibility of producing inverse Compton (IC) and/or pion decay emission is scarce. In Figure \ref{fig:deeppointing} we show a wide field of view intensity map with a long integration time where the \agl\ is not visible and its position is marked with a white cross.


\begin{figure}[ht!]
\resizebox{\hsize}{!}{\includegraphics{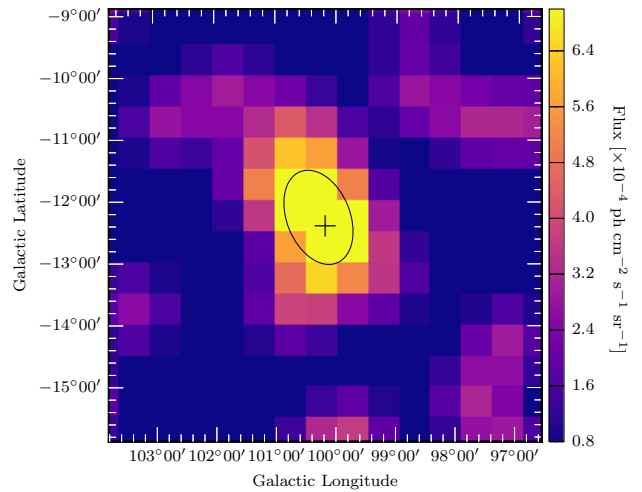}}
\caption{\agile\ 2-day integration intensity map corresponding to data published in \cite{Lucarelli10} at date 2010--07--25 (MJD 55402). Black ellipse represents the 95\% c.l. containment of gamma-ray flux and black cross marks the nominal position of the Be star \mwc. Color-map represents 3-pixels kernel gaussian smoothed number of counts, with a pixel size of 0\grp5. \label{fig:mapAtel}}
\end{figure}


\subsubsection{Search for gamma-ray transient emission}

On the 27$^{th}$ of July 2010 the \agile\ Team reported  transient gamma-ray activity detected from an unidentified source at Galactic Coordinates $(l,b)=($100\grp0, -12\grp2$)$ during the period 2010-07-25 01:00 -- 2010-07-26 23:30 UT \citep{Lucarelli10}. A refined maximum likelihood analysis of this transient event yields a 5.3~$\sigma$ detection with a flux of $F(E>100 \rm{MeV})= (2.0 \pm 0.7)\times 10^{-6}$ cm$^{-2}$ s$^{-1}$. Figure \ref{fig:mapAtel} shows the intensity map above 100 MeV for this event.

After the gamma-ray detection in 2010 and the discovery of \mwc\ as a Be/BH binary system in a position compatible with \agl, a search for transient emission at other epochs has been performed. A blind search multi-source likelihood analysis around the position of \agl\ for energies above 100~MeV has been conducted in 2-day bins with no overlap between them, yielding a total of 10 gamma-ray transient or flaring events at a position compatible, within their 95\% position uncertainties, with the one reported in \cite{Lucarelli10}, and with the position of \mwc. These transient events spread from 2007 up to 2013, including both pointing and spinning observations and the details about each one are displayed in Table \ref{tab:detections}. In this Table we list the peak position of the transient emission, in galactic coordinates, the date and time of the starting and ending of the transient events, the flux and the significance as $\sqrt{TS}$. In Figure \ref{fig:lc} we show the flux of these detections plotted over time, in MJD units, toghether with the significance of each one . A total of 151 and 758 2-day bins were analyzed for pointing and spinning observations, respectively.

Given that the significance of most of the events found is only a bit above 3$\sigma$ we computed the probability of this detections to be by chance using the procedure explained in \cite{bulgarelli2012b}. With this method we computed the probability of having $k$ detections -consistent with the position of \agl, and hence with \mwc- with a significance $\sqrt{TS} \ge \sqrt{h}$ with $N$ trials, which is given by:

\begin{equation*}
P(N,k) = 1 - \sum_{j=0}^{k-1} \dbinom{N\,}{j\,} \; p^{j} (1-p)^{N-j}
\end{equation*}

where $p$ is the $p$-value corresponding to $h$ which, for a detection significance $\sqrt{TS} \ge 3$ for a source below the galactic plane whose counterpart might be known, is $p = 1.3\times10^{-3}$. For the $N=909$ 2-day bins in which we searched for transient events the probability of the 10 detections being by chance is $P(909,10) = 6.8\times10^{-7}$, which translates into 4.97 Gaussian standard deviations. Given this probability and the fact that the detections are always compatible with the position of \mwc, we think that it might be reasonable to consider the 10 detected events as true transient gamma-ray emission.

\begin{figure}[t!]
\resizebox{\hsize}{!}{\includegraphics{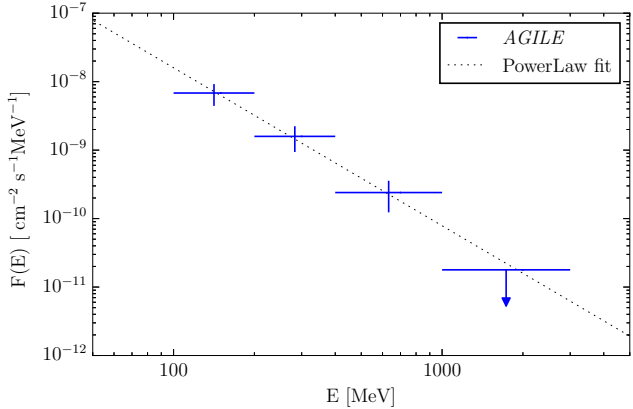}}
\caption{Photon spectrum between 100~MeV and 3~GeV of \mwc\ as detected by \agile/GRID by integrating all flaring episodes in pointing and spinning mode. Dotted line represents the best powerlaw fit, with a photon index $\Gamma = 2.3 \pm 0.2$. \label{fig:spectrum}}
\end{figure}

\begin{figure}[t!]
\resizebox{\hsize}{!}{\includegraphics{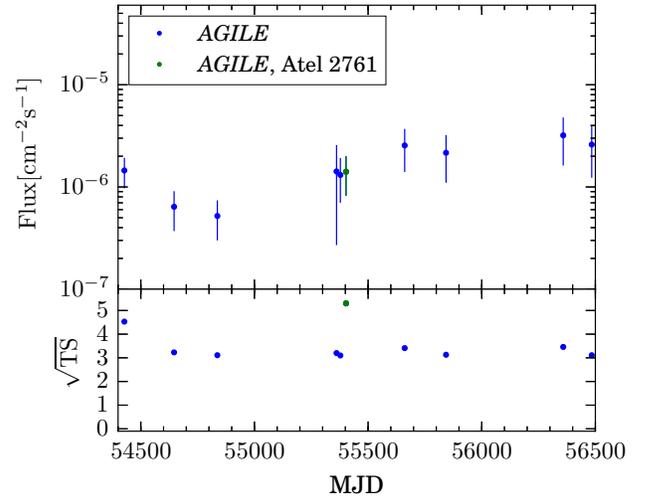}}
\caption{\agile\ detections plotted over time. The green data point corresponds to the event published in \cite{Lucarelli10}. \label{fig:lc}}
\end{figure}

With this result in mind, a stacked analysis of the 10 detections was carried out and yielded a best position for our source of interest of $(l, b) = ($100\grp37, -12\grp39$) \pm$ 0\grp35. The spectrum, that is displayed in Figure \ref{fig:spectrum}, was binned into four channels: 100--200~MeV, 200--400~MeV, 400--1000~MeV and 1000--3000~MeV, and it was fitted with a powerlaw obtaining a photon index $\Gamma = 2.3 \pm 0.2$. The overall significance for this stacked analysis is $\sqrt{TS}=8.9$.


\begin{table*}[t!]
      \begin{center}
    \caption{\agile\ gamma-ray transient detections around the position of \mwc. \label{tab:detections}}        
    \begin{tabular}{cccc}
    \hline
    \hline
t$_{start}$ & t$_{end}$ & Flux & \multirow{2}{*}{$\sqrt{TS}$} \\
$[UT]$ & $[UT]$ & $[\times 10^{-6}$ cm$^{-2}$ s$^{-1}]$ & \\

\hline

2007-11-23 UT00:00:00 & 2007-11-24 UT00:00:00 & 1.5 $\pm$ 0.5 & 4.5 \\
2008-06-28 UT00:00:00 & 2008-06-30 UT00:00:00 & 0.6 $\pm$ 0.3 & 3.2 \\
2009-01-04 UT00:00:00 & 2009-01-07 UT00:00:00 & 0.5 $\pm$ 0.2 & 3.1 \\
\hline
2010-06-13 UT00:00:00 & 2010-06-14 UT00:00:00 & 1.4 $\pm$ 1.1 & 3.2 \\
2010-06-30 UT00:00:00 & 2010-07-02 UT00:00:00 & 1.3 $\pm$ 0.6 & 3.1 \\
2010-07-25 UT00:00:00 & 2010-07-27 UT00:00:00 & 1.4 $\pm$ 0.6 & 5.3 \\
2011-04-09 UT00:00:00 & 2011-04-11 UT00:00:00 & 2.2 $\pm$ 1.1 & 3.1 \\
2011-10-08 UT00:00:00 & 2011-10-10 UT00:00:00 & 2.5 $\pm$ 1.1 & 3.4 \\
2013-03-07 UT00:00:00 & 2013-03-08 UT09:00:00 & 2.6 $\pm$ 1.4 & 3.1 \\
2013-07-10 UT00:00:00 & 2013-07-12 UT00:00:00 & 3.2 $\pm$ 1.6 & 3.5 \\

\hline
    \end{tabular}
    \end{center}
\end{table*}


\subsubsection{Search for gamma-ray periodic emission}

Since the gamma-ray transient AGL~J2241+4454 has been proposed to be the counterpart of \mwc\ we have conducted a search for periodicity in the gamma-ray flares detected by \agile. We have folded the transient detections with the period of \mwc, in accordance with \cite{Casares14} ephemeris, which are displayed in Figure \ref{fig:orbit}. The detected flares span between phase 0 and 0.8 approximately, leaving a small gap without any gamma-ray detection. There is no evident periodic behaviour in the gamma-ray emission from \agl\ that might be associated to \mwc. This kind of emission, however, was not expected, since the orbital parameters of the system seem to point at a very small eccentricity \citep{Casares14}.


\subsection{\fermi\ data analysis}

For the purpose of this work, we used the Science Tools provided by the \textit{Fermi} satellite team\footnote{{\tt http://fermi.gsfc.nasa.gov}} on the Pass8 data around the position of \agl. The version of the Science Tools used was {\tt v9r33p0} with the P8R2\_TRANSIENT010\_V6 instrument response function (IRF). The reader is referred to  \textit{Fermi} instrumental publications for further details about IRFs and other calibration details \citep{irfs}.

We have adopted the current Galactic diffuse emission model ({\tt gll\_iem\_v06.fits}) in a likelihood analysis and {\tt iso\_P8R2\_SOURCE\_V6\_v06.txt} as the isotropic model, and the \fermi\ 3rd point source catalog {\tt gll\_psc\_v16.fit} \citep{Fermi3fgl} has been used\footnote{{\tt http://fermi.gsfc.nasa.gov/ssc/data/access/}}. In the modelisation of the data, the galactic background and diffuse components remained fixed. We selected  Pass8 FRONT and BACK transient class events with energies between 0.1 and 300 GeV. Among them, we limited the reconstructed zenith angle to be less than 105$^{\circ}$ to greatly reduce gamma rays coming from the limb of the Earth's atmosphere. We selected the good time intervals of the observations by excluding events that were taken while the instrument rocking angle was larger than 50$^{\circ}$. 

In the model for our source we  used a power-law model for \mwc\ and used the {\tt make3FGLxml.py} tool to obtain a model for the sources within 25$^{\circ}$ region of inrerest (ROI). To analyze the data we used the user contributed package {\it Enrico}\footnote{{\tt https://github.com/gammapy/enrico/}}. 

We divided each analisys in two steps: in the first one we leave all parameters of all the sources within a 10$^{\circ}$ ROI free, while the sources outside this ROI up to 25$^{\circ}$ have their parameters fixed. Then we run a likelihood analysis using the Minuit optimizer, which determines the spectral-fit parameters, and obtain a fit for all these sources. In the second step, we fix all the parameters of the sources in our model to the fitted values, except for our source of interest, and run again the likelyhood analysis with the Newminuit optimizer to obtain a refined fit. At all times, the central target source \mwc\ kept all its spectral parameters free.

A systematic search in the \fermi\ data was carried out searching for transient emission on a 2-day time integration bins retrieving non significant events. The time intervals of the 8 \agile\ flares during which \fermi\ was already operational were also searched and no detections arose, yielding upper limits at the level of a few $10^{-7}$ ph cm$^{-2}$ s$^{-1}$. We stacked these 8 time intervals in order to be more sensitive and performed an analysis that resulted also in a non detection of the source, yielding an upper limit at the level of $3\times10^{-7}$ ph cm$^{-2}$ s$^{-1}$ above 100~MeV. We have also performed a search for periodic emission in the \fermi\ data by folding the data with the orbital period of 60.37 days of \mwc. For this purpose we divided the orbit in 8 bins to ensure having enough statistics and repeated the analysis procedure as explained. No detection arose at any particular orbital phase and upper limits were obtained at the level of $\sim 2\times10^{-9}$ ph cm$^{-2}$ s$^{-1}$. In the next Section we discuss the implications of these results.

\subsubsection{Comparison between \agile\ and \fermi\ data}

Given the distinct results found by both gamma-ray telescopes we wanted to investigate why this had happened, as it is not the first time that one event is seen only by one of these telescopes (see for instance \citealt{Sabatini2013}). It is important to note that the detectability of transient emission for a specific source depends on the effective area (larger by a factor 5--10 in \fermi\ compared to \agile\ depending upon energy in the range 100~MeV--1~GeV, and decreasing with off-axis angle for both instruments), on the effective time on source (depending on the observation mode, which is different in the two telescopes) and marginally on the spectral energy distribution of the emission (with \agile\ optimized in the 100-400 MeV channel and \fermi\ optimized for higher energies).

\begin{figure}[t!]
    \includegraphics[width=\linewidth]{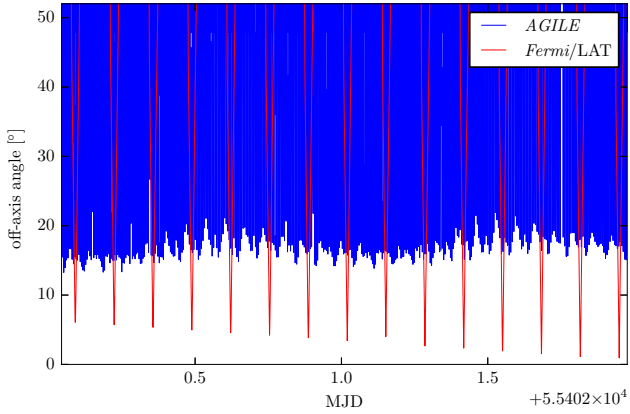}
    \caption{Time-evolution of the \mwc\ off-axis angles, as observed by \agile\ and \fermi\ during the time interval reported by \cite{Lucarelli10}. \label{fig:check}}

\end{figure}

For example, Figure \ref{fig:check} shows the time evolution of the source off-axis angle - i.e., the angular distance between \mwc\ and the field of view (FoV) center - for each instrument, during the event detected by \agile\ on 2010-07-25 \citep{Lucarelli10}. We can note that - most of the time - the source is inside the \agile-GRID FoV at small off-axis angles, whereas it transits at large off-axis angles (or outside the FoV) in \fermi: \mwc\ is observed at angular distance lower than 50$^\circ$ in ~41\% of the total time for \agile\ and ~25\% for \fermi. It is important to remark that, at high values of the off-axis angle ($> 50^\circ$), the sensitivity of \fermi\ is poor respect to the nominal on-axis value (~50\% smaller).


\begin{figure}[t!]
\resizebox{\hsize}{!}{\includegraphics{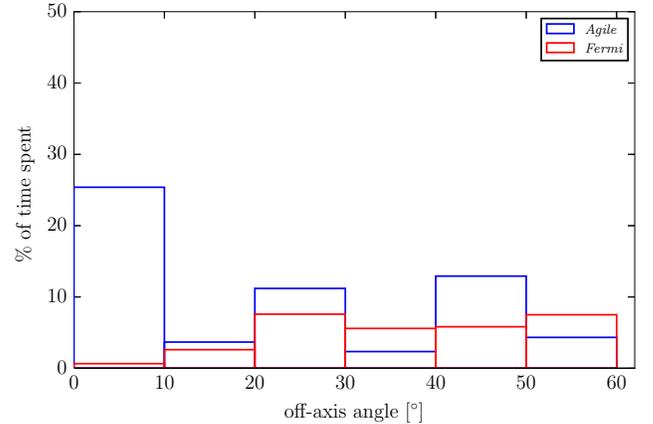}}
\caption{\agile\ and \fermi\ histograms for the off-axis angle of \mwc\ during all the detected transient events. \label{fig:histogram}}
\end{figure}

The other time intervals of transient emission detected by \agile\ are also marginally exposed by \fermi\ and Figure \ref{fig:histogram} shows the fraction of time during which the source has been observed at a given interval of the off-axis angle by \agile\ and \fermi\, during all the events in Table~\ref{tab:detections}. \agile\ has a much better angular exposition of the source with respect to \fermi. The low exposed \fermi\ observations, related to a reduced visibility of the source, provides a flux UL of 2.6 $\times 10^{-8}$~cm$^{-2}$~s$^{-1}$ for the stacked analysis of all the events in Table~\ref{tab:detections}, and typical ULs of the order of $\sim 9\times10^{-8}$~cm$^{-2}$~s$^{-1}$ for a 2-day integration time. Thus, these values are not consistent with the \agile\ results. However, these ULs are related to average gamma-ray emission during the 2-day time intervals. Nevertheless, short radio flares, possibly related to the binary system activity, have been recently observed (Dzib et al., 2015). Therefore in the following we discuss the hypothesis that the flare duration is shorter than the integrated time.

A soft spectrum flare with a short duration of about $\sim$3-4h could have been seen by only one telescope if the arrival times of the (few) incident photons are within the good time intervals (GTI) of that telescope. \cite{alexander2015} studied the discrepance between \agile\ and \fermi. In their Fig. 4, they show that, during the time intervals of higher \agile\ count rates, \fermi\ was not in a GTI data-taking, and possibly lost most of the photons detected by \agile. In order to discuss these cases, we carried out simulations of short transient emission with {\tt gtobssim} (\fermi\ Science Tools). We simulated a source with the same flux as the average flux seen by \agile\ in the 2010 event (2.0$\times 10^{-6}$ ph cm$^{-2}$ s$^{-1}$) and with the same spectral index ($\Gamma = 2.3$). The same attitude spacecraft file for real data was used in the simulation in order to use the exact positioning and GTIs of the satellite. To do such a test, we assumed that the flare observed by AGILE lasted for a few hours and the flux that we obtained is diluted within the 2-day integration time that was used in the analysis. Hence, we simulated a source with a higher flux than the observed, but lasting only 4 to 6 hours out of the total 2-day simulated time interval, in order to have a realistic scenario. The results of this simulations with the \fermi\ Science Tools show that a short lived flare lasting for 4 or 6~h would not be seen by \fermi, obtaining a significance of ~2 and ~3 sigma, and upper limits at the level of $1.1\times10^{-7}$ and $2.0\times10^{-7}$ ph cm$^{-2}$ s$^{-1}$, respectively.

\section{Discussion}

The discovery of \agl\ by \agile\ has triggered a series of studies that lead to the discovery of the first Be/BH binary system, \mwc. However, the actual nature and counterpart of \agl\ remains still uncertain. Usual counterparts of flaring or transient gamma-ray events are blazars, novae and gamma-ray or X-ray binaries. However, the portion of the sky where \agl\ was found seems to be poorly populated with these kind of objects. The most suitable and also interesting candidate is still \mwc.

\begin{figure}[t!]
\resizebox{\hsize}{!}{\includegraphics{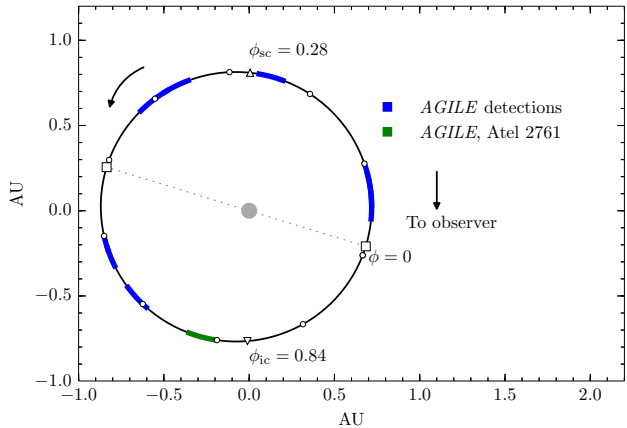}}
\caption{Sketch of the orbit of \mwc\ according to the ephemeris from \cite{Casares14}. The \agile\ detected transient events are plotted on top. Squares mark the position of the periastron and the apastron whereas triangles mark the position of inferior and superior conjunctions. The orbit of the BH, the size of the star and the duration of the \agile\ observations are scaled to the real values. \label{fig:orbit}}
\end{figure}

Among the sources that can be found within the error box of \agl, we considered \mwc\ as the main candidate to be the counterpart of the gamma-ray activity. Within this assumption, we have plotted the gamma-ray transient events folded with the period of 60.37 days of this binary system (see Figure \ref{fig:orbit}).  We found that the transient emission is distributed along the orbit with no preferred orbital phase for the gamma-ray activity, i.e., there is no clustering of the detections at any particular phase. At some of these orbital phases the compact object is behind the Be star (see Figure \ref{fig:orbit}). Other binary systems like Cygnus X-3 have been found to emit in gamma-rays \citep{Tavani2009,Abdo2009Sci}. The modulated gamma-ray emission shows a bright maximum at the superior conjunction (where the  compact object is behind the donor star). This behavior can be explained by assuming a simple leptonic scenario (anisotropic efficiency of the IC scattering, \citealt{2010MNRAS.404L..55D}), or a hadronic scenario with an anisotropic or clumpy wind \citep{2014ApJ...780...29S}. 

The orbit of the binary system is almost circular, with an eccentricity of 0.10$\pm$0.04 \citep{Casares14}. With such a low eccentricity, there is almost no difference between the periastron and apastron passages, and thus the interaction between the compact object and the donor star is expected to be almost steady in time. A possible source of variability within the system might be instabilities in the accretion disk around the BH and/or variations in the strong equatorial wind of the Be star. \cite{Hubert1998} observed that rapid variability in the optical emission in Be stars occured more often in early type Be stars, as it is the case of \mwc. If this variability occurs in this binary and it affects the accretion regime onto the BH, this might explain the gamma-ray transient events. 

\mwc\ has been recently classified as a HMXB \citep{munaradrover14} and during the X-ray observation in 2013 the source exhibited a very low luminosity, compatible with a quiescent state. The recent VLA radio detections by \cite{Dzib15} showed variable radio emission with a peak at a flux density of $14.2\pm2.9~\mu$Jy for a single $\sim$2h observation at orbital phase $\sim$0.5. During other radio observations the source went into udetectable levels in radio, below a few $\mu$Jy. Only by stacking this other observations the source was detected again with a flux density of $\sim 4\mu$Jy. This proofs the variability within the binary system and supports the possible variable gamma-ray emission as well, since it is related to changes in the accretion regime that could give rise to an increase in the acceleration of particles up to relativistic energies.

\cite{aleksic2015} reported upper limits to the possible VHE emission of \mwc\ observed with the MAGIC Telescopes. These upper limits are at the level of $2.0\times10^{-12}$ cm$^{-2}$ s$^{-1}$ and are not compatible with an extrapolation of the spectrum reported here up to TeV energies. This, however, does not exclude the possibility of VHE emission from \mwc, since the MAGIC observations were carried out during a quiescent X-ray state of the binary, and VHE emission should not be expected during this state.

The accretion onto a BH in a system like \mwc\ might be explained by an advection dominated accretion flow (ADAF) \citep{Narayan1994}. In this scenario, a very low accretion rate onto the compact object occurs, giving rise to the very low X-ray emission. At radio wavelengths the source would display either a very low flux or it would not emit at all, as it has been seen in \cite{Dzib15}, which reinforces the ADAF scenario. The radio emission could be interpreted as synchrotron radiation arising from a comptonized environment of electrons near the BH. These population of electrons could be part of a corona and/or be a non-already colimated jet.

In Cygnus X-3 the transient gamma-ray events occur during soft X-ray states or during transition states that preceed major radio flares \citep{Piano2012}, while in Cygnus X-1 gamma-ray flares are more sporadic and were observed during hard states \citep{Albert2007, Sabatini2010} or during hard-to-soft transitions \citep{Sabatini2013}. In the case of \mwc\, the source seems to be always in a quiescent state and the lack of simultaneous data during gamma-ray events prevent us to discuss the gamma-ray flares in relation to possible state transitions. 

If coming from the binary, the gamma-ray flares could be produced within the comptonized corona that surrounds the BH. In this case, relativistic electrons scatter soft photons either from the accretion disk or from the compation Be star. However, special conditions need to be fulfilled in order to give rise to the transient gamma-ray episodes. 
If coming from \mwc, the gamma-ray luminosity of the 2010 event would be $L_{\rm \gamma} = (9.0 \pm 5.2)\times 10^{35}$ erg cm$^{-2}$ s$^{-1}$, for a distance to the system of 2.6 $\pm$ 0.6 kpc.

\section{Conclusions}

In July 2010 \agile\ detected the transient source \agl\ which triggered the studies that revealed the first Be/BH binary system. Further searches in the \agile\ data revealed that 9 other flares from the same location occured between 2008 and 2013 with energies above 100~MeV. By stacking all the detected events a spectral characterization of the source has been possible yielding a fit to a powerlaw with photon index $\Gamma=2.3\pm 0.2$ between 100~MeV and 3~GeV. The field around \agl\ does not contain many possible counterpart sources. In fact we consider only two: RX~J2243.1+4441, a radio galaxy with a possible FR-II type classification and thus less probable to be a gamma-ray emitter; and \mwc, the above mentioned HMXB containing a Be star and a BH. We consider the latter as the most suitable source to produce the observed gamma-ray emission, given its X-ray and radio characteristics. In order to do a more complete study, we confronted the \agile\ data with \fermi\ data, performing the same study with both telescopes. However, \fermi\ data does not show evidence of the observed gamma-ray emission. Nevertheless, we think that it might be due to differences in the way both telescopes observe, as well as due to different energy-dependent sensitivity. Simultaneous radio to gamma-ray data in a wide temporal coverage is needed to unveil the identity of \agl.


\acknowledgements

AGILE is an ASI space mission developed with programmatic support by INAF and INFN. We acknowledge partial support through the ASI grant no. I/028/12/2. This research made use of Enrico, a community-developed Python package to simplify Fermi-LAT analysis \citep{enrico13}

\newpage

\newpage
\begin{appendix}

\section{The \agl\ field}\label{ap:field}
In this appendix we show the nearby X-ray sources that might be the counterpart of \agl\ as well as the gamma-ray sources taken into account in the likelihood analysis which might be sources of contamination. Within the error box of the \agile\ detection we find the two main candidates RX~J2243.1+4441 and MWC~656. As explained in Section \ref{sec:field}, there are two nearby gamma-ray sources that have been taken into account in the data analysis: 3FGL~J2247.8+4413 which is a possible blazar located at $\sim9^\circ$ from \agl, and 3FGL~J2247.8+4413 a possible BL Lac object located at $\sim1^\circ$ from \agl.

\begin{figure}[h!]
    \begin{minipage}{.5\textwidth}

        \includegraphics[width=\linewidth]{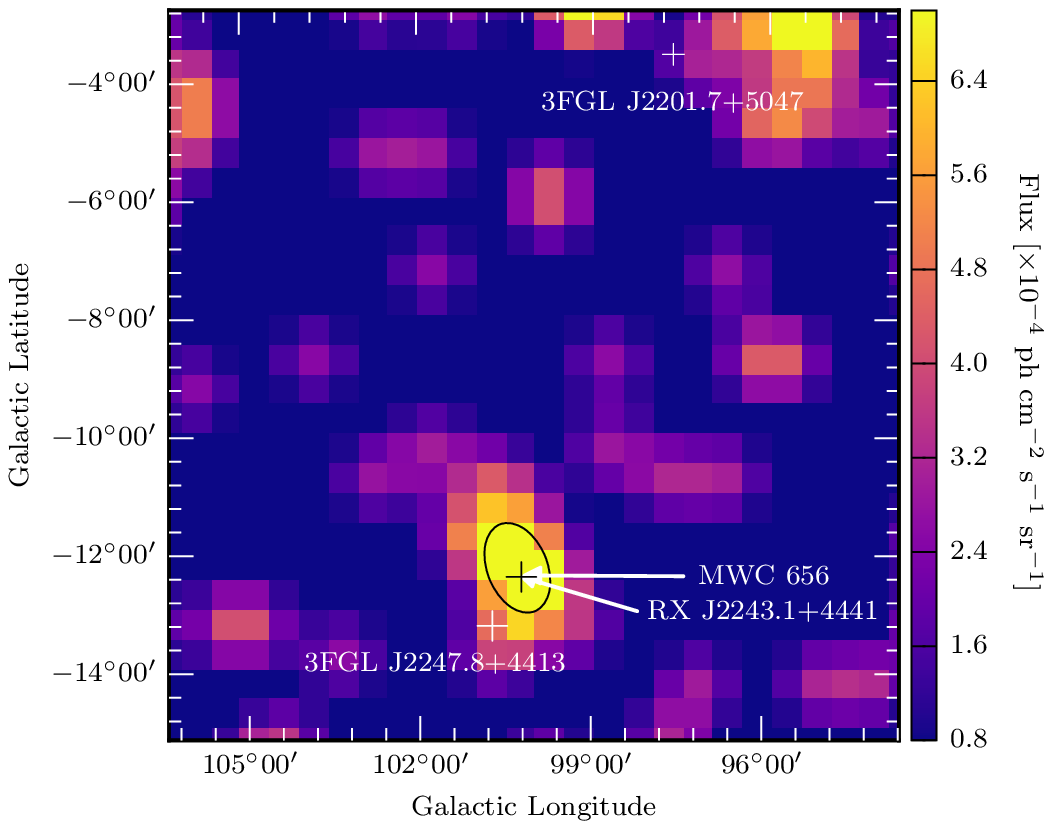}
        
    \end{minipage}
    \begin{minipage}{0.5\textwidth}

        \includegraphics[width=\linewidth]{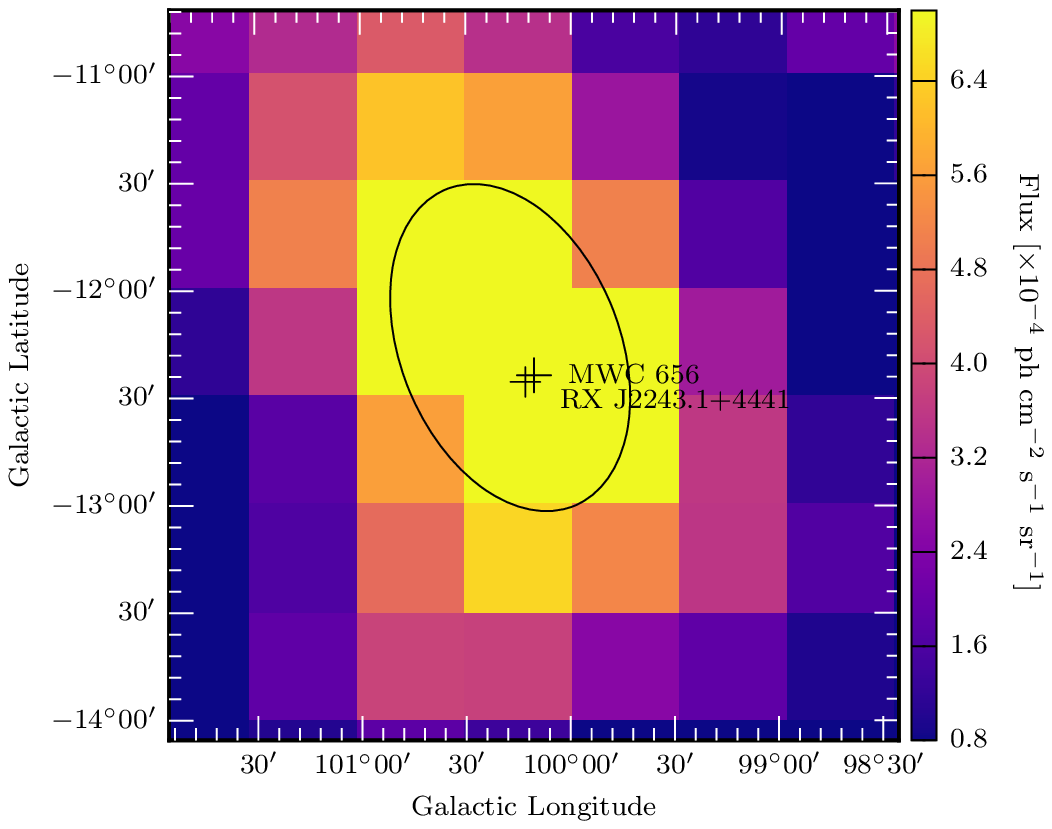}
        \label{fig:nearbysources}
    \end{minipage}
    \caption{\agile\ intensity map corresponding to the 2010 detected event. {\it Left}: The black ellipse corresponds to the 95\% error box of the \agile\ detection; within the error box there are \mwc\ and RX~J2243.1+4441, which are very nearby. The source 3FGL~J2201.7+5047 lies in the upper part of the map, while a possible BL Lac object, 3FGL~J2247.8+4413, is located on the lower left part of the map. {\it Right}: zoom in of the \agl\ region and the sources inside the error box.}
\end{figure}

\begin{table*}[t!]
      \begin{center}
    \caption{\agl\ nearby sources \label{tab:field}}        
    \begin{tabular}{lcccccl}
    \hline
    \hline
Name & RA         & Dec                                & l          & b          & Distance to& Notes \\
     & [h:m:s] & [$^\circ$:$^\prime$:$^{\prime\prime}$] & [$^\circ$] & [$^\circ$]& \agl\ [$^\circ$] &       \\
\hline
MWC~656              & 22:42:57.3 & $+$44:43:18.3 & 100.1755 & -12.3985 & 0.19 & Be/BH binary system \\
RX~J2243.1+4441      & 22:43:14.7 & $+$44:42:31.0 & 100.2152 & -12.4350 & 0.16 & Possible FR-II Radio Galaxy \\
3FGL~J2247.8+4413    & 22:47:53.2 & $+$44:13:15.5 & 100.7236 & -13.2583 & 0.93 & Possible BL Lac \\
3FGL~J2201.7+5047    & 22:01:44.2 & $+$50:48:00.0 & 97.6417  & -03.5460 & 9.24 & Gamma-ray source\\
\hline
    \end{tabular}
    \end{center}
\end{table*}

\end{appendix}

\end{document}